\documentclass[12pt, a4paper, onecolumn]{article}

\usepackage{fullpage}
\usepackage{hyperref}

\usepackage[english]{babel}
\usepackage{amsmath}         
\usepackage{amssymb}         

\usepackage{graphicx}
\usepackage{url}           

\long\def\symbolfootnote[#1]#2{
\begingroup
	\def\thefootnote{\fnsymbol{footnote}}\footnote[#1]{#2}
\endgroup}

\begin{document}

\title{Learning to Rank for Expert Search in Digital Libraries of Academic Publications}

\author{Catarina Moreira\\ \small \texttt{catarina.p.moreira@ist.utl.pt}\\
\and
P\'{a}vel Calado\\ \small \texttt{pavel.calado@ist.utl.pt}
\and
Bruno Martins\\ \small \texttt{bruno.g.martins@ist.utl.pt}
\and
\\Instituto Superior T\'{e}cnico, INESC-ID\\ Av. Professor Cavaco Silva, 2744-016 Porto Salvo, Portugal\\  
\\ \small The original publication is available at: Progress in Artificial Intelligence, Springer\\
 \small \text{\url{http://link.springer.com/chapter/10.1007\%2F978-3-642-24769-9\_32}}
}




\date{}

\maketitle

\begin{abstract}
The task of expert finding has been getting increasing attention in information retrieval literature. However, the current state-of-the-art is still lacking in principled approaches for combining different sources of evidence in an optimal way. This paper explores the usage of learning to rank methods as a principled approach for combining multiple estimators of expertise, derived from the textual contents, from the graph-structure with the citation patterns for the community of experts, and from profile information about the experts. Experiments made over a dataset of academic publications, for the area of Computer Science, attest for the adequacy of the proposed approaches.
\end{abstract}

\symbolfootnote[0]{This work was partially supported by the ICP Competitiveness and Innovation Framework Program of the European Commission, through the European Digital Mathematics Library (EuDML) project -- \url{http://www.eudml.eu/}}

\section{Introduction}

The automatic search for knowledgeable people in the scope of specific user communities, with basis on documents describing people's activities, is an information retrieval problem that has been receiving increasing attention~\cite{Pavel09Search}. Usually referred to as \emph{expert finding}, the task involves taking a short user query as input, denoting a topic of expertise, and returning a list of people sorted by their level of expertise in what concerns the query topic. 

Several effective approaches for finding experts have been proposed, exploring different retrieval models and different sources of evidence for estimating expertise. However, the current state-of-the-art is still lacking in principled approaches for optimally combining the multiple sources of evidence that can be used to estimate expertise. In traditional information retrieval tasks such as ad-hoc retrieval, there has been an increasing interest on the usage of machine learning methods for building retrieval formulas capable of estimating relevance for query-document pairs~\cite{Liu09Learning}. The general idea is to use hand-labeled data (e.g., document collections containing relevance judgments for specific sets of queries, or information regarding user-clicks aggregated over query logs) to train ranking models, this way leveraging on data to combine the different estimators of relevance in an optimal way. However, few previous works have specifically addressed the usage of learning to rank approaches in the task of expert finding.
	
This paper explores the usage of learning to rank methods in the expert finding task, specifically combining a large pool of estimators for expertise. These include estimators derived from the textual similarity between documents and queries, from the graph-structure with the citation patterns for the community of experts, and from profile information about the experts. We have built a prototype expert finding system using learning to rank techniques, and evaluated it on an academic publication dataset from the Computer Science domain.

The rest of this paper is organized as follows: Section 2 presents the main concepts and related works. Section 3 presents the learning to rank approaches used in our experiments. Section 4 introduces the multiple features upon which we leverage for estimating expertise. Section 5 presents the experimental evaluation of the proposed methods, detailing the dataset and the evaluation metrics, as well as the obtained results. Finally, Section 6 presents our conclusions and points directions for future work.

\section{Concepts and Related Work}

Serdyukov and Macdonald have surveyed the most important concepts and representative previous works in the expert finding task~\cite{Pavel09Search,Macdonald08Voting}. Two of the most popular and well-performing types of methods are the profile-centric and the document-centric approaches~\cite{Craswell06Overview,Soboroff07Overview}. Profile-centric approaches build an expert profile as a pseudo document, by aggregating text segments relevant to the expert~\cite{Balog06Formal}. These profiles of experts are latter indexed and used to support the search for experts on a topic. 
Document-centric approaches are typically based on traditional document retrieval techniques, using the documents directly. In a probabilistic approach to the problem, the first step is to estimate the conditional probability $p(q|d)$ of the query topic $q$ given a document $d$. Assuming that the terms co-occurring with an expert can be used to describe him, $p(q|d)$ can be used to weight the co-occurrence evidence of experts with $q$ in documents. The conditional probability $p(c|q)$ of an expert candidate $c$ given a query $q$ can then be estimated by aggregating all the evidences in all the documents where $c$ and $q$ co-occur. Experimental results show that document-centric approaches usually outperform profile-centric approaches~\cite{Soboroff07Overview}. 

Many different authors have proposed sophisticated probabilistic retrieval models, specific to the expert finding task, with basis on the document-centric approach~\cite{Balog06Formal,Petkova07Proximity,Pavel09Search}. For instance Cao et al. proposed a two-stage language model combining document relevance and co-occurrence between experts and query terms~\cite{Cao06Research}. Fang and Zhai derived a generative probabilistic model from the probabilistic ranking principle and extend it with query expansion and non-uniform candidate priors~\cite{Fang07Probabilistic}. Zhu et al. proposed a multiple window based approach for integrating multiple levels of associations between experts and query topics in expert finding~\cite{Zhu07Open}. More recently, Zhu et al. proposed a unified language model integrating many document features for expert finding~\cite{Zhu08Modeling}. Although the above models are capable of employing different types of associations among query terms, documents and experts, they mostly ignore other important sources of evidence, such as the importance of individual documents, or the co-citation patterns between experts available from citation graphs. In this paper, we offer a principled approach for combining a much larger set of expertise estimates.

In the Scientometrics community, the evaluation of the scientific output of a scientist has also attracted significant interest due to the importance of obtaining unbiased and fair criteria. Most of the existing methods are based on metrics such as the total number of authored papers or the total number of citations. A comprehensive description of many of these metrics can be found in~\cite{Sidiropoulos05citation,Sidiropoulos06generalized}. Simple and elegant indexes, such as the Hirsch index, calculate how broad the research work of a scientist is, accounting for both productivity and impact. Graph centrality metrics inspired on PageRank, calculated over citation or co-authorship graphs, have also been extensively used~\cite{Liu05authorship}. In the context of academic expert search systems, these metrics can easily be used as query-independent estimators of expertise, in much the same way as PageRank is used in the case of Web information retrieval systems.

For combining the multiple sources of expertise, we propose to leverage on previous works concerning the subject of learning to rank for information retrieval (L2R4IR). Tie-Yan Liu presented a good survey on the subject~\cite{Liu09Learning}, categorizing the previously proposed algorithms into three groups, according to their input representation and optimization objectives:

\begin{itemize}
\item {\bf Pointwise approach} - L2R4IR is seen as either a regression or a classification problem. Given feature vectors of each single document from the data for the input space, the relevance degree of each of those individual documents is predicted with scoring functions which can sort all documents and produce the final ranked list.

\item {\bf Pairwise approach} - L2R4IR is seen as a binary classification problem for document pairs, since the relevance degree can be regarded as a binary value which tells which document ordering is better for a given pair of documents. Given feature vectors of pairs of documents from the data for the input space, the relevance degree of each of those documents can be predicted with scoring functions which try to minimize the average number of misclassified document pairs. Several different pairwise methods have been proposed, including SVM$rank$~\cite{Joachims06rank}.

\item {\bf Listwise approach} - L2R4IR is addressed in a way that takes into account an entire set of documents, associated with a query, as instances. These methods train a ranking function through the minimization of a listwise loss function defined on the predicted list and the ground truth list. Given feature vectors of a list of documents of the data for the input space, the relevance degree of each of those documents can be predicted with scoring functions which try to directly optimize the value of a particular information retrieval evaluation metric, averaged over all queries in the training data~\cite{Liu09Learning}. Several different listwise methods have also been proposed, including SVM$map$~\cite{Yue07Support}.
\end{itemize}

In this paper, we made experiments with the application of representative learning to rank algorithms from the pairwise and the listwise approaches, namely the SVM{\it rank} and the SVM{\it map} algorithms, in a task of expert finding within digital libraries of academic publications.

\section{Learning to Rank Experts}

\begin{figure*}[ht!]
\resizebox{\columnwidth}{!} {
\includegraphics[scale=0.33]{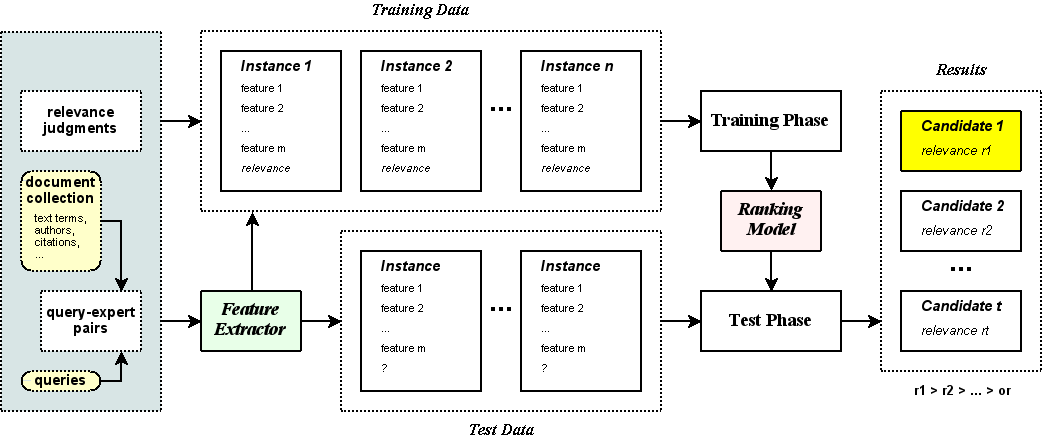}
}
\caption{The general procedure of learning to rank for expert search.}
\label{f1}
\end{figure*}

In this paper, we follow a general approach which is common to most supervised learning to rank methods, consisting of two separate steps, namely training and testing. Figure~\ref{f1} provides an illustration.

Given a set of queries $Q = \{q_1,\ldots, q_{|Q|}\}$ and a collection of experts $E = \{e_1,\ldots, e_{|E|}\}$, each associated with specific documents describing the topics of expertise, a training corpus for learning to rank is created as a set of query-expert pairs, each $(q_i, e_j) \in Q \times E$, upon which a relevance judgment indicating the match between $q_i$ and $e_j$ is assigned by a labeler. This relevance judgment can be a binary label, e.g., relevant or non-relevant, or an ordinal rating indicating relevance, e.g., definitely relevant, possibly relevant, or non-relevant. For each instance $(q_i, e_j)$, a feature extractor produces a vector of features that describe the match between $q_i$ and $e_j$. Features can range from classical IR estimators computed from the documents associated with the experts (e.g., term frequency, inverse document frequency, BM25, etc.) to link-based features computed from networks encoding relations between the experts in $E$ (e.g., PageRank). The inputs to the learning algorithm comprise training instances, their feature vectors, and the corresponding relevance judgments. The output is a ranking function, $f$, where $f(q_i, e_j)$ is supposed to either give the true relevance judgment for $(q_i, e_j)$, or produce a ranking score for $e_j$ so that when sorting experts according to these scores the more relevant ones appear on the top of the ranked list.

During the training process, the learning algorithm attempts to learn a ranking function capable of sorting experts in a way that optimizes a particular bound on an information retrieval performance measure (e.g., Mean Average Precision). In the test phase, the learned ranking function is applied to determine the relevance between each expert $e_j$ in $E$ and a new query $q$. In this paper, we experimented with the following learning to rank algorithms:

\begin{itemize}
\item SVM$rank$~\cite{Joachims06rank} : This pairwise method builds a ranking model in the form of a linear scoring function, i.e. $f(x)=w^Tx$, through the formalism of Support Vector Machines (SVMs). The idea is to minimize the following objective function over a set of $n$ training queries $\{q_i\}_{i=1}^n$, their associated pairs of experts $(x_u^{(i)},x_v^{(i)})$ and the corresponding relevance judgment $y_{u,v}^{(i)}$ over each pair of experts (i.e., pairwise preferences resulting from a conversion from the ordered relevance judgments over the query-expert pairs):
\begin{equation}
\begin{split}
\min \frac{1}{2}||w||^2 + C \sum_{i=1}^n \sum_{u,v:y_{u,v}^{(i)}} \xi_{u,v}^{(i)} \\
\text{s.t.~} w^T(x_u^{(i)} - x_v^{(i)}) >= 1 - \xi_{u,v}^{(i)} {\text{ , if~~}} y_{u,v}^{(i)} = 1, \xi_{u,v}^{(i)} >= 0 {\text{ ,~~}} i=1, \dots,n
\end{split}
\end{equation}

Differently from standard SVMs, the loss function in SVM$rank$ is a hinge loss defined over document pairs. The margin term $\frac{1}{2}||w||^2$ controls the complexity of the pairwise ranking model $w$. 
The method introduces slack variables, $\xi_{u,v}^{(i)}$, (i.e., a variable that is added to an optimization constraint to turn an inequality into an equality where a linear combination of variables is less than or equal to a given constant), which measure the degree of misclassification of the datum $x_i$. The coefficient $C$ affects the trade-off between model complexity and the proportion of non-separable samples. If it is too large, we have a high penalty for non-separable points and we may store many support vectors and overfit. If it is too small, we may have underfitting.
The objective function is increased by a function which penalizes non-zero $\xi_{u,v}^{(i)}$, and the optimization becomes a trade off between a large margin, and a small error penalty.

\item SVM$map$~\cite{Yue07Support} : This listwise method builds a ranking model through the formalism of structured Support Vector Machines~\cite{Tsochantaridis05Structured}, attempting to optimize the metric of Average Precision (AP). Suppose $x=\{x_j\}_{j=1}^m$ is the set of all the experts associated with a training query $q$, and $y_{u,v}^{(i)}$ represents the corresponding ground truth labels. Any incorrect label for $x$ is represented as $y^c$. The SVM$map$ approach can be formalized as follows, where AP is used in the constraints of the structured SVM optimization problem.
\begin{center}
\begin{equation}
\begin{split}
\min \frac{1}{2}||w||^2 + \frac{C}{n} \sum_{i=1}^n \xi^{(i)}\\
\text{s.t.~} \forall y^{c(i)} \neq y^{(i)} , w^T \Psi(y^{(i)},x^{(i)}) >= w^T \Psi(y^{c(i)},x^{(i)}) + 1 - AP(y^{c(i)}) - \xi^{(i)}
\end{split}
\end{equation}
\end{center}
In the constraints, $\Psi$ is called the joint feature map, whose definition is:
\begin{equation}
\begin{split}
\Psi(y,x) = \sum_{u,v:y_u=1, y_v=0} (x_u - x_v) \\
\Psi(y^c,x) = \sum_{u,v:y_u=1, y_v=0} (x_u^c, y_v^c)(x_u - x_v)
\end{split}
\end{equation}
Since there are an exponential number of incorrect labels for the documents, it is a big challenge to directly solve the optimization problem involving an exponential number of constraints for each query. The formalism of structured SVMs efficiently tackles this issue by maintaining a working set with those constraints with the largest violation:
\begin{equation}
{\text Violation} \triangleq 1 - AP(y^c) + w^T\Psi(y^c,x)
\end{equation}
\end{itemize}
The survey by Tie-Yan Liu discusses the above methods in more detail~\cite{Liu09Learning}.

\section{Features for Estimating Expertise}

The considered set of features for estimating the expertise of a researcher towards a given query can be divided into three groups, namely textual features, profile features and graph features. The textual features are similar to those used in standard text retrieval systems and also in previous learning to rank experiments (e.g., TF-IDF and BM25 scores). The profile similarity features correspond to importance estimates for the authors, derived from their profile information (e.g., number of papers published). Finally, the graph features correspond to importance and relevance estimates computed from the author co-authorship and co-citation graphs.

\subsection{Features Based on Textual Similarity}

Similarly to previous expert finding proposals based on document-centric approaches, we also use textual similarity between the query and the contents of the documents to build estimates of expertise. In the domain of academic digital libraries, the associations between documents and experts can easily be obtained from the authorship information associated to the publications. For each topic-expert pair, we used the Okapi BM25 document-scoring function, to compute the textual similarity features. Okapi BM25 is a state-of-the-art IR ranking mechanism composed of several simpler scoring functions with different parameters and components (e.g., term frequency and inverse document frequency). It can be computed through the formula shown in Equation~\ref{eq:bm25}, where $Terms(q)$ represents the set of terms from query {\it q}, ${\it Freq(i,d)}$ is the number of occurrences of term {\it i} in document $d$, $|d|$ is the number of terms in document $d$, and $\mathcal{A}$ is the average length of the documents in the collection. The values given to the parameters $k_1$ and $b$ were 1.2 and 0.75 respectively. Most previous IR experiments use these default values for the $k_1$ and $b$ parameters.

\begin{equation} \label{eq:bm25}
\begin{split}
{BM25(q,d)} = \sum_{i \in Terms(q)}\log \left( \frac{N-Freq(i)+0.5}{Freq(i)+0.5} \right) \times\\ \frac{(k_1+1) \times \frac{Freq({i,d})}{|d|}} {\frac{Freq({i,d})}{|d|} + k_1 \times (1 - b + b \times \frac{|d|}{\mathcal{A}})}
\end{split}
\end{equation} 
We also experimented with other textual features commonly used in ad-hoc IR systems, such as {\it Term Frequency} and {\it Inverse Document Frequency}.

Term Frequency (TF) corresponds to the number of times that each individual term in the query occurs in all the documents associated with the author. Equation~\ref{eq:tf} describes the TF formula, where $Terms(q)$ represents the set of terms from query {\it q}, $Docs(a)$ is the set of documents having {\it a} as author, $Freq(i, d_{j})$ is the number of occurrences of term {\it i} in document $d_{j}$ and $\left|d_{j}\right|$ represents the number of terms in document $d_{j}$.
	\begin{equation}
TF_{q,a} =  \sum_{j \in Docs(a)} \sum_{i \in Terms(q)} \frac{Freq({i,d_j})}{|d_j|}
\label{eq:tf}
\end{equation}

The Inverse Document Frequency (IDF) is the sum of the values for the inverse document frequency of each query term and is given by Equation~\ref{eq:idf}. In this formula, $|D|$ is the size of the document collection and $f_{i, D}$ corresponds to the number of documents in the collection where the $i_{th}$ query term occurs.
		
\begin{equation}
	\label{eq:idf}
	IDF_q = \sum_{i \in Terms(q)} \log \frac{|D|}{f_{i, D}}
\end{equation}

Other features used were the number of unique authors associated with documents containing the query topics, the range of years since the first and last publications of the author containing the query terms, and the document length, in terms of the number of words, for all the publications associated to the author.

In the computation of these textual features, we considered two different textual streams from the documents, namely (i) a stream consisting of the titles, and (ii) a stream using the abstracts of the articles. 

\subsection{Features Based on Profile Information}

We also considered a set of profile features related to the amount of published materials associated with authors, generally taking the assumption that highly prolific authors are more likely to be considered experts. Most of the features based on profile information are query independent, meaning that they have the same value for different queries. The considered set of profile features are based on the temporal interval between the first and the last publications, the average number of papers and articles per year, and the number of publications in conferences and in journals with and without the query topics in their contents.

\subsection{Features Based on Graphs Co-citation and Co-authorship}

Scientific impact metrics computed over scholarly networks, encoding co-citation and co-authorship information, can offer effective approaches for estimating the importance of the contributions of particular publications, publication venues, or individual authors. Thus, we considered a set of features that estimate expertise with basis on co-citation and co-authorship information. The features considered are divided in two sets, namely (i) citation counts and (ii) academic indexes. In what regards citation counts, we used the total, the average and the maximum number of citations of papers containing the query topics, the average number of citations per year of the papers associated with an author and the total number of unique collaborators which worked with an author. On what regards academic impact indexes, we used the following features:

\begin{itemize}
\item {\bf Hirsch index} of the author and of the author's institution, measuring both the scientific productivity and the apparent scientific impact~\cite{Hirsch05Index}. An author/institution has an Hirsch index of $h$ if $h$ of his $N_p$ papers have at least $h$ citations each, and the other $(N_p - h)$ papers have at most $h$ citations each. Authors with a high Hirsch index, or authors associated with institutions with a high Hirsch index, are more likely to be considered experts.
\item The {\bf $h$-$b$-index}, which extends the Hirsch index for evaluating the impact of scientific topics in general~\cite{Banks06extension}. In our case, the scientific topic is given by the query terms and thus the query has an $h$-$b$-index of $i$ if $i$ of the $N_p$ papers containing the query terms in the title or abstract have at least $i$ citations each, and the other $(N_p - i)$ papers have at most $i$ citations each.

\item {\bf Contemporary Hirsch index} of the author, which adds an age-related weighting to each cited article, giving less weight to older articles~\cite{Antonis06Generalized}. A researcher has a contemporary Hirsch index $h^c$ if $h^c$ of his $N_p$ articles get a score of $S^c(i) >= h^c$ each, and the rest $(N_p - h^c)$ articles get a score of $S^c(i) <= h^c$. For an article $i$, the score $S^c(i)$ is defined as:
\begin{equation}
S^c(i) = \gamma * (Year(now) - Year(i) + 1)^{-\delta} * |CitationsTo(i)|
\end{equation}
The $\gamma$ and $\delta$ parameters are set to $4$ and $1$, respectively, meaning that the citations for an article published during the current year account four times, the citations for an article published 4 years ago account only one time, the citations for an article published 6 years ago account $4/6$ times, and so on.

\item {\bf Trend Hirsch index}~\cite{Antonis06Generalized} for the author, which assigns to each citation an exponentially decaying weight according to the age of the citation, this way estimating the impact of a researcher's work in a particular time instance. A researcher has a trend Hirsch index $h^t$ if $h^t$ of his $N_p$ articles get a score of $S^t(i) >= h^t$ each, and the rest $(N_p - h^t)$ articles get a score of $S^t(i) <= h^t$. For an article $i$, the score $S^t(i)$ is defined as:
\begin{equation}
S^t(i) = \gamma * \sum_{\forall x \in C(i)} (Year(now) - Year(x) + 1)^{-\delta}
\end{equation}
The $\gamma$ and $\delta$ parameters are set to $4$ and $1$, respectively.

\item {\bf Individual Hirsch index} of the author, computed by dividing the value of the standard Hirsch index by the average number of authors in the articles that contribute to the Hirsch index of the author, in order to reduce the effects of frequent co-authorship with influential authors~\cite{Batista06Possible}.

\item The {\bf $a$-index} of the author/institution, measuring the magnitude of the most influential articles. For an author or institution with an Hirsch index of $h$ that has a total of $N_{c,tot}$ citations toward his papers, we say that he has an $a$-index of $a = N_{c,tot} / h^2$.

\item The {\bf $g$-index} of the author/institution, also quantifying scientific productivity with basis on the publication record~\cite{Egghe06Theory}. Given a set of articles associated with the author/institution, ranked in decreasing order of the number of citations that they received, the g-index is the unique largest number $g$ such that the top $g$ articles received on average at least $g$ citations.

\item The {\bf $e$-index} of the author~\cite{ZhangEIndex} which represents the excess amount of citations of an author. The motivation behind this index is that we can complement the $h$-index by taking into account these excess amounts of citations which are ignored by the $h$-index.
The $e$-index is given by the Equation~\ref{eq:e-index}, where $cit_j$ are the citations received by the $j_th$ paper and $h$ is the $h$-index.
\begin{equation}
e = \sum_{j=1}^{h} \sqrt{cit_j-h^2}
\label{eq:e-index}
\end{equation}
\end{itemize}

Besides the above features, and following the ideas of Chen et al.~\cite{Chen07Finding}, we also considered a set of graph features that estimate the influence of individual authors using PageRank, a well-known graph linkage analysis algorithm that was introduced by the Google search engine. 

PageRank assigns a numerical weighting to each element of a linked set of objects (e.g., hyperlinked Web documents or articles in a citation network) with the purpose of measuring its relative importance within the set. The PageRank value of a node is defined recursively and depends on the number and PageRank scores of all other nodes that link to it (i.e., the incoming links). A node that is linked to by many nodes with high PageRank receives a high rank itself. 

Formally, given a graph with $N$ nodes $i=1,2,\cdots,N$, with $L$ directed links that represent references from an initial node to a target node with weights $\alpha=1,2,\cdots,L$, the PageRank $Pr_i$ for the $i$th node is defined by:

\begin{equation}
Pr_i = \frac{0.5}{N} + 0.5 \sum_{j \in inlinks(L,i)} \frac{\alpha_j Pr_j}{outlinks(L,j)}
\end{equation}

In the formula, the sum is over the neighboring nodes $j$ in which a link points to node $i$. The first term represents the random jump in the graph, giving a uniform injection of probability into all nodes in the graph. The second term describes the propagation of probability corresponding to a random walk, in which a value at node $j$ propagates to node $i$ with probability $\frac{\alpha_j Pr_j}{outlinks(L,j)}$. 

The features that we considered correspond to the sum and average of the PageRank values associated to the papers of the author that contain the query terms, computed over a directed graph representing citations between papers. Each citation link in the graph is given a score of $1/N$, where $N$ represents the number of authors in the paper. Authors with high PageRank scores are more likely to be considered experts.

\section{Experimental Validation}

The main hypothesis behind this work is that learning to rank approaches can be effectively used in the context of expert search tasks, in order to combine different estimators of relevance in a principled way, this way improving over the current state-of-the art. To validate this hypothesis, we have built a prototype expert search system, reusing existing implementations of state-of-the-art learning to rank algorithms, namely the SVM$rank$\footnote{\url{http://www.cs.cornell.edu/people/tj/svm_light/svm_rank.html}} implementation by Thorsten Joachims ~\cite{Joachims06rank} and the SVM$map$\footnote{\url{http://projects.yisongyue.com/svmmap/}} implementation by Yue et al~\cite{Yue07Support}.

 We implemented the methods responsible for computing the features listed in the previous section, using {\it Microsoft SQL Server 2008} (e.g., the full-text search capabilities for computing the textual similarity features) and several existing Java software packages (e.g., the LAW\footnote{\url{http://law.dsi.unimi.it/software.php}} package for computing PageRank). 

The validation of the prototype required a sufficiently large repository of textual contents describing the expertise of individuals within a specific area. In this work, we used a dataset for evaluating expert search in the Computer Science research domain, corresponding to an enriched version of the DBLP\footnote{\url{http://www.arnetminer.org/citation}} database made available through the Arnetminer project.

DBLP data has been used in several previous experiments regarding citation analysis~\cite{Sidiropoulos05citation,Sidiropoulos06generalized} and expert search~\cite{Deng08Formal}. It is a large dataset covering both journal and conference publications for the computer science domain, and where substantial effort has been put into the problem of author identity resolution, i.e., references to the same persons possibly with different names. Table~\ref{t1} provides a statistical characterization of the DBLP dataset.
\begin{table*}[t!]
\resizebox{\columnwidth}{!} {
\begin{tabular}{l c}
  
  Property																																		& Value								\\
  \hline
  Total Authors 																															& ~~~~1 033 050~~~~		\\
  Total Publications 																													& ~~~~1 632 440~~~~ 	\\
  Total Publications containing Abstract																			& ~~~~653 514~~~~ 		\\
  Total Papers Published in Conferences~~~~~~~~~~~~~~~~~~~~~~~~~~~~~~~~~~~~~~ & ~~~~606 953~~~~			\\
  Total Papers Published in Journals 																					& ~~~~436 065~~~~ 		\\
  Total Number of Citations Links 																						& ~~~~2 327 450~~~~ 	\\
  \hline
\end{tabular}
}
\scriptsize
\caption{Statistical characterization of the DBLP dataset used in our experiments}
\label{t1}
\end{table*}

To train and validate the different learning to rank methods, we also needed a set of queries with the corresponding author relevance judgments. For the Computer Science domain, we used the relevant judgments provided by Arnetminer\footnote{\url{http://arnetminer.org/lab-datasets/expertfinding/}} which have already been used in other expert finding experiments~\cite{yang09bole}.

The Arnetminer dataset comprises a set of 13 query topics from the Computer Science domain, each associated to a list of expert authors. In order to add negative relevance judgments (i.e., complement the dataset with unimportant authors for each of the query topics), we searched the dataset with the keywords associated to each topic, retrieving the top $n/2$ authors according to the BM25 metric and retrieving $n/2$ authors randomly selected from the dataset, where $n$ corresponds to the number of expert authors associated to each particular topic. This way, we obtained twice the relevant judgments provided by Arnetminer, ending up with 2794 records for all 13 queries. Table~\ref{judgments} shows the distribution for the number of experts associated to each topic, as provided by Arnetminer.
\begin{table*}[ht!]
\resizebox{\columnwidth}{!} {
\begin{tabular}{l c l c}

~~{\bf Query Topics}			& ~~{\bf Rel. Authors}		&~~{\bf Query Topics} 				&~~{\bf Rel. Authors}	\\
\hline

 ~~Boosting (B)									& ~~46									 				&	~~Natural Language (NL)					& ~~41						\\
 ~~Computer Vision (CV)					&	~~176									 				&	~~Neural Networks	 (NN)					& ~~103						\\
 ~~Cryptography (C)							&	~~148									 				& ~~Ontology				 (O)					& ~~47						\\
 ~~Data Mining (DM)							&	~~318									 				& ~~Planning				 (P)					& ~~23						\\
 ~~Information Extraction (IE)	&	~~20									 				& ~~Semantic Web		 (SW)					& ~~326						\\
 ~~Intelligent Agents	(IA)			&	~~30									 				& ~~Support Vector Machines (SVM)	& ~~85						\\
 ~~Machine Learning	(ML)				& ~~34									 				& 																&								\\
\hline
\end{tabular}
}
\caption{Characterization of the Arnetminer dataset of Computer Science experts.}
\label{judgments}
\end{table*}

The test collection was used in a leave-one-out cross-validation methodology, in which different experiments used 9 different queries to train a ranking model, which was then evaluated over the remaining queries. The averaged results from the four  different cross-validation experiments are finally used as the evaluation result. To measure the quality of the results produced by the different learning to rank algorithms, we used two different performance metrics, namely the Precision@k (P@k) and the Mean Average Precision (MAP).

Precision at rank $k$ is used when a user wishes only to look at the first $k$ retrieved domain experts. The precision is calculated at that rank position through Equation~\ref{eq:PrecisionRank}.
\begin{equation}
P@k=\frac{r\left(k\right)}{k}
\label{eq:PrecisionRank}
\end{equation}
In the formula, $r(k)$ is the number of relevant authors retrieved in the top {\it k} positions. $P@k$ only considers the top-ranking experts as relevant and computes the fraction of such experts in the top-$k$ elements of the ranked list.

The Mean of the Average Precision over test queries is defined as the mean over the precision scores for all retrieved relevant experts. For each query $r$, the Average Precision (AP) is given by:

\begin{equation}
AP[r] = \frac{\sum_{k=1}^n P@k[r] \times I\{ g_{r_k} = \max(g) \}}{\sum_{k=1}^n I\{ g_{r_k} = \max(g) \}}  
\end{equation}

As before, $n$ is the number of experts associated with query $q$ and $g_{rk}$ is the relevance grade for author $k$ in relation to the query $r$. In the case of our datasets, $\max(g) = 1$ (i.e., we have 2 different grades for relevance, 0 or 1).

Table~\ref{t2} presents the obtained results over the DBLP dataset. The obtained results attest for the adequacy of both learning to rank approaches, showing that SVM$rank$ and SVM$map$ achieve a similar performance, with SVM$rank$ slightly outperforming SVM$map$ in our experiments in terms of MAP.

\begin{table*}[ht!]
\resizebox{\columnwidth}{!} {
\begin{tabular}{ l c c c c c }

~~~~~~~~~~~~~~~~~~~~~~~ &~~~~P@5~~~~ 		& ~~~~P@10~~~~ 	& ~~~~P@15~~~~ 	& ~~~~P@20~~~~ 	& ~~~~MAP~~~~ 	\\
\hline
~~SVM$rank$          		& 0.9333							& {\bf 0.9104}	& {\bf 0.8848}	& 0.8698	& {\bf 0.8150}  \\
~~SVM$map$       				& {\bf 0.9458}  			& 0.8979  			& 0.8778 				& {\bf 0.8721}  			& 0.8131  			\\
\hline
\end{tabular}
}
\caption{Results of the SVM$map$ and SVM$rank$ methods.}
\label{t2} 
\end{table*}

In a separate experiment, we attempted to measure the impact of the different types of ranking features on the quality of the results. Using the best performing learning to rank algorithm, SVM$rank$, we separately measured the results obtained by ranking models that considered (i) only the textual similarity features, (ii) only the profile features, (iii) only the graph features, (iv) only a representative graph feature, namely the h-$b$-index, (v) textual similarity and profile features, (vi) textual similarity and graph features and (vii) profile and graph features. Table~\ref{t3} shows the obtained results, also presenting the previous results reported by Yang et al.~\cite{yang09bole} over the same dataset, as well as the results obtained by the h-$b$-index bibliographic index.

\begin{table*}[ht!]
\resizebox{\columnwidth}{!} {
\begin{tabular}{ l c c c c c }
~~~~~~~ 															 & ~~~~P@5~~~~	& ~~~~P@10~~~~	& ~~~~P@15~~~~ & ~~~~P@20~~~~	& ~~~~MAP~~~~	\\
\hline
Text Similarity + Profile + Graph~~  		&{\bf 0.9333}	& {\bf 0.9104}	& {\bf 0.8848} & {\bf 0.8698}	& {\bf 0.8150} 			\\
Text Similarity + Profile							 	& 	0.6917		 	& 0.6583  			& 0.6861  		 & 0.6552				& 0.6601 			\\
Text Similarity + Graph					   	 		&  0.9250 			& 0.8934				& 0.8167			 & 0.7896				& 0.7677 			\\
Profile + Graph										 	 		&  0.8667	 		& 0.8250				& 0.8273			 & 0.8125				& 0.7943	 \\
Text Similarity        									&  0.7042			& 0.6646				& 0.6597			 & 0.6511				& 0.6569				\\
Profile       												 	&  0.7500			& 0.7646				& 0.7389			 & 0.7313				& 0.7464     		\\
Graph        										 		 		&  0.8750			& 0.8438				& 0.8181			 & 0.8021				& 0.7846   			\\
\hline
h-$b$-Index														 	& 0.7385       & 0.7077    		& 0.6821    	 & 0.6700	      & 0.6053 \\
\hline
Expert Finding (Yang et al.)~\cite{yang09bole} &	0.5500			& 0.6000				& 0.6333			 &~~--	& 0.6356 \\
\hline

\end{tabular}
}
\caption{The results obtained with different sets of features and comparison with other approaches.}
\label{t3}
\end{table*}

As we can see, the set with the combination of all features has the best results. The results also show that, individually, textual similarity features have the poorest results. This means that considering only textual evidence provided by query topics, together with article's titles and abstracts, may not be enough to determine if some authors are experts or not, and that indeed the information provided by citation and co-authorship patterns can help in expert retrieval. Finally, the results show that the different combinations of all features proposed in this paper outperform the previously proposed learning to rank approach for expert finding made by Yang et al.~\cite{yang09bole}

\begin{figure*}[ht!]
\resizebox{\columnwidth}{!} {
\includegraphics[scale=0.25]{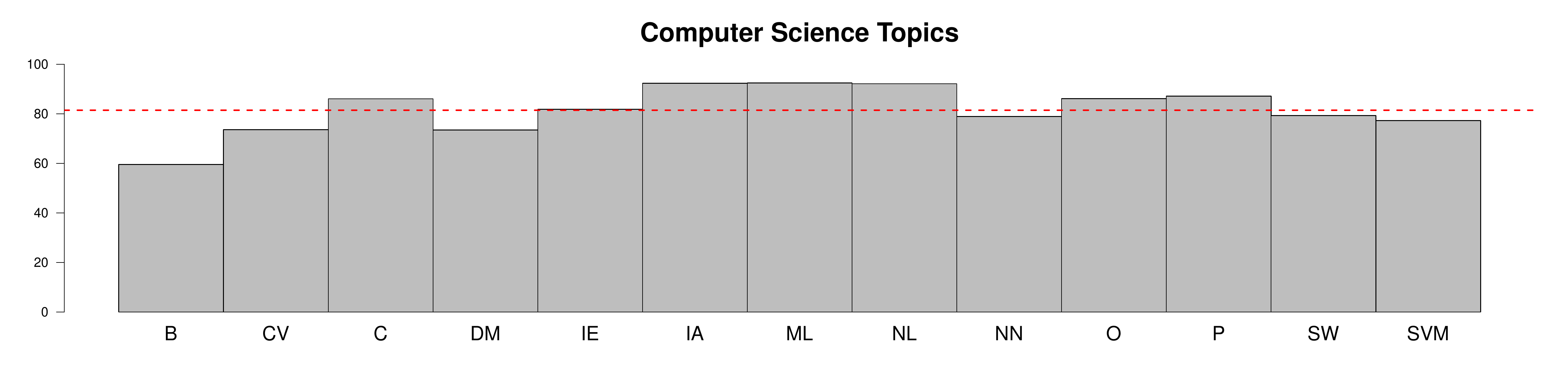}
}
\caption{Average precision over the different query topics.}
\label{f2}
\end{figure*}

Figure~\ref{f2} plots the obtained average precision in each of the individual query topics for the best performing approach, namely SVM$rank$ with the combination of all features. The figure presents the query topics in the same order as they are given in Table~\ref{judgments}. The horizontal dashed line corresponds to the MAP obtained in the same experiment. The results show that there are only slightly variations in performance for the different queries.

Finally, Table~\ref{tab:people} shows the top five people which were returned by the system for four different queries, corresponding to the best and worst results in terms of the P@5 metric. The system performed well for the queries Neural Networks, Machine Learning and Support Vector Machines (SVMs). Although these are very related topics, the system managed to distinguish between them and still identify the relevant experts in these areas correctly.  However, worse results were returned for the query Boosting.  These poor results can be explained by the absence of the query topics in the titles and abstracts of the publications of authors working in the area. We realized that the authors which were judged as relevant, and therefore considered experts, did not have too many query topics present in their publication's titles or abstracts, leading to misclassifications. 
\begin{table*}[ht!]
\resizebox{\columnwidth}{!} {
\begin{tabular}{c c c c c |c }
\hline
  \multicolumn{3}{ c }{{\bf Best Results}} 						  			& {\bf Worst Results}\\   
	  \hline
  ~~~~{\bf Neural Networks}~~~~& ~~~{\bf Machine Learning}~~~ & {\bf SVMs}& ~~~{\bf Boosting}~~~	\\
  \hline
  
 	Geoffrey E. Hinton					 & Robert E. Schapire	 					& Thorsten Joachims					& J. Ross Quinlan 					\\	
	Erkki Oja				 						 & Vladimir Vapnik							& Robert E. Schapire					& B. Han 	\\	
 	Yann LeCun									 & Thomas G. Dietterich					& Vladimir Vapnik							& W. Shireen 		\\	
 	Thomas G. Dietterich				 & Michael I. Jordan						& Christopher J. C. Burges		& L. Carlos de Freitas 		\\	
  Michael I. Jordan						 & Manfred K. Warmuth						& Tomaso Poggio								& Robert E. Schapire 	 		\\
 
  \hline
\end{tabular}
}
\caption{Top five people returned by the system for four different queries.}
\label{tab:people}
\end{table*}

\section{Conclusions}

This paper explored the usage of learning to rank methods in the context of expert searching within digital libraries of academic publications. We argue that learning to rank provides a sound approach for combining multiple estimators of expertise, derived from the textual contents, from the graph-structure of the community of experts, and from expert profile information. Experiments on datasets of academic publications show very good results in terms of P@5 and MAP, attesting for the adequacy of the proposed approaches.

Despite the interesting results, there are also many ideas for future work. Recent advancements in the area of learning to rank for information retrieval are, for instance, concerned with query-dependent ranking (i.e., using different ranking models according to the type of queries being issued) and it would be interesting to test these techniques in expert searching tasks.

Our approach to the expert finding problem can also be generalized to any type of entity search. The introduction of Entity Ranking Track in INEX 2007, with basis on a {\it Wikipedia} dataset, provides a good platform for general entity search evaluation~\cite{Vries08Overview}. For future work, it would be interesting to experiment with learning to rank methods, similar to the ones proposed in this paper, over the more general entity search problem.

\bibliographystyle{plain}

\end{document}